\documentclass[article,11pt]{amsart}
\usepackage{amsmath, amssymb, amsthm, amsfonts, amsgen, stmaryrd}
\usepackage[english]{babel}
\usepackage{fullpage}
\usepackage[centertags]{amsmath}
\usepackage{indentfirst}
\usepackage{fancyhdr}
\usepackage[dvips]{graphicx}
\usepackage{psfrag}
\usepackage{enumerate}
\numberwithin{equation}{section}
\usepackage[latin1]{inputenc}
\usepackage{amstext}
\usepackage{array}

\usepackage{color}

\newtheorem{theo}{Theorem}[section]

\newtheorem{rema}[theo]{Remark}
\newtheorem{example}{Example}[section]

\newcommand{\R}{{\mathbb{R}}}
\newcommand{\Z}{{\mathbb{Z}}}

\newcommand{\C}{\mathbb{C}}
\newcommand{\T}{\mathcal{T}}

\newcommand{\dd}{\mathrm{d}}
\newcommand{\del}{\partial}

\newcommand{\nn}{\nonumber}
\newcommand{\beq}{\begin{equation}}
\newcommand{\eeq}{\end{equation}}

\newcommand{\D}{\mathcal{D}}

\newcommand{\g}{\mathfrak{g}}

\newcommand{\A}{\mathcal{A}}

\newcommand{\F}{\mathcal{F}}
\newcommand{\DD}{{\mathcal Der}}
\newcommand{\cS}{\mathcal S}
\newcommand{\cO}{\mathcal O}
\newcommand{\cP}{\mathcal P}
\newcommand{\cA}{\mathcal A}
\newcommand{\cL}{\mathcal{L}}
\newcommand{\cB}{\mathcal{B}}
\newcommand{\cH}{\mathcal{H}}
\newcommand{\cF}{\mathcal{F}}
\newcommand{\cM}{\mathcal{M}}

\title{Abstract dynamical systems: \\ remarks on symmetries and reduction}
\date{25 august 2020}
\author{Giuseppe Marmo}
\address{Dipartimento di Fisica ``E. Pancini'', Universit\`a di Napoli Federico II; INFN - Sezione di Napoli, Via Cintia - 80126 Napoli, Italy }
\email{marmo@na.infn.it}
\author{Alessandro Zampini}
\address{Dipartimento di Matematica ed Applicazioni ``R. Caccioppoli'', Universit\`a di Napoli Federico II; INFN - Sezione di Napoli, Via Cintia - 80126 Napoli, Italy }
\email{azampini@na.infn.it}

\begin{document}

\thispagestyle{empty}
\begin{abstract}
We review how an algebraic formulation for the dynamics of a physical system allows to describe a reduction procedure for both classical and quantum evolutions. 
\end{abstract}

\maketitle
\tableofcontents

\section{An introduction}
\label{sec:intro}

One of the most important achievements of last century mathematics is the discovery that it is possible to study mathematical structures \emph{per se}, without any specific realisation or representation. An interesting example for this is provided by the  notion of Lie group, which can be studied without referring to any group of transformations. The transformation-point-of-view has originated the Erlangen programme, elaborated by F.~Klein with the help of S.~Lie. 
Within physics, this path was developed by Dirac in the formulation of dynamics, classical and quantum, as well as of field theory. He indeed formulated a \emph{principle of analogy}. 
In this note we would like to take such point of view and describe an abstract notion of system (evolutionary or dynamical), its symmetries and the reduction procedures.

Refined  notions of \emph{abstract} dynamical system have been analysed in order to  unify  the description of the  time evolution of a physical system 
when formulated in terms of ordinary differential equations (both linear and non linear), or of integral equations, or of partial differential equations (see \cite{walker-book80, infinite-book84, slemrod70}). 

Following the quantum ideology (i.e. following the guiding idea that the geometrical formulation of  classical mechanics is related to  a geometrical formulation of  quantum mechanics, and that many mathematical properties of both classical and quantum dynamics can be stated in purely algebraic terms, postponing the physical identification of the variables to the stage when the condition of dealing with $c$-number or $q$-number is imposed \cite{dirac-qft}) we mean to 
describe physical systems in terms of an (abstract) algebra and of its states (that is elements of its dual space), and to study the time evolution when formulated   via one parameter groups of automorphisms, infinitesimally generated by derivations of the algebras. Given such a description, in section \ref{sec3} we describe an algebraic scheme for the reduction of a classical dynamics, and begin to  generalise it in section \ref{sec4} to the quantum setting.

\subsection*{Acknowledgments} This paper originates from the talk that one of us (G.M.) was invited to deliver at the conference \emph{Non commutative manifolds and their symmetries}, dedicated to Giovanni Landi on the occasion of his 60th birthday. We dedicate this paper to him, whose PhD thesis extensively dealt with algebraic differential calculus for physical theories. It is a pleasure for us to thank the organisers of the conference, and to acknowledge the support of the INFN, the INDAM, and the Santander UC3M Excellence Chair Programme 2019/2020.

\section{An algebraic description of a dynamical system}
\label{sec2}

\subsection{Observables and states}\label{ssec2.1} Any mathematical formulation for the dynamics of a physical system requires to identify a set $\cS$ of \emph{states}, which represent the maximal information about the system, a set $\cO$ of \emph{observables}, i.e. measurable quantities for the system, a \emph{pairing}, i.e. a map
$$
\mu\,:\,\cO\,\times\,\cS\quad\rightarrow\quad\cP 
$$
with $\cP$ the set of probability measures on $\R$. Given a state $\rho\in\cS$ and $A\in\cO$, the quantity 
$$
0\,\leq\,\mu(A,\rho)(\Delta)\,\leq\,1
$$
provides the probability that the measurement of $A$ while the system is in the state $\rho$ gives a result in $\Delta$, an element in the Borel $\sigma$-algebra over $\R$. Treating systems with an arbitrary number of constituents requires to have a \emph{composition law}, under which to define composite systems out of simple ones. 
 
A somehow minimal setting allowing to formalise the above notions is to consider the  set of observables $\cO$  as the real subspace  of a normed unital $*$-algebra $\cA$ over the complex field\footnote{The use of complex structures in physics has been found convenient to describe, for instance, interference phenomena and seems unavoidable in quantum field theory in order to describe creation and annihilation of particles. By a normed $*$-algebra we mean an associative algebra $\A$ with a norm with respect to which all the algebraic operations are continuous.} $\C$. The norm on $\A$ induces a topology on $\A^*$, and states are the positive normalised elements in $\A^*$: the pairing is naturally given in terms of the duality between $\A$ and $\A^*$, the composition of elementary systems by taking the tensor product of the corresponding algebras. 

\subsection{Derivations and dynamics}\label{ssec2.2}We formulate  a  dynamics for a physical systems associated to an algebra $\A$, that is a rule of time evolution for the observables (or, alternatively, for the states)  as a one parameter group $\Phi_t$ (with $t\in\R$) of suitable automorphisms on $\A$ or dually on  the states $\cS\subset\cA$. Notice that we are not considering the evolution in terms of a dynamical semigroup of a quantum system interacting with an environment, nor in terms of implicit equations. This allows to say that infinitesimal generators for the evolution are given by\footnote{To be definite, we recall here that $\Phi_t$ is a one-parameter group of automorphisms for $\A$ if $\Phi_t$ is a $*$-automorphism for $A$ for any $t\in\R$,  it is norm continuous (i.e. $\lim_{t\to0}\|\Phi_t(a)-a\|=0$ for any $a\in\A$) and  fulfills  the composition property that $\Phi_{t+t'}=\Phi_t\circ\Phi_{t'}$ with $\Phi_0=\mathbb{I}$. On any $*$-algebra, a linear operator $\delta\,:\,D(\delta)\to\A$ from a dense $*$-subalgebra $D(\delta)\subseteq\A$ (the domain of $\delta$) such that $\delta(a^*)=\delta(a)^*$ and $\delta(aa')=a\delta(a')+(\delta(a))a'$ is called a $*$-derivation.  If $\Phi_t$ is a one parameter group of automorphisms for $\A$, its infinitesimal generator is the $*$-derivation defined as the linear operator $\delta(a)=\lim_{t\to0}(\Phi_t(a)-a)/t$ on the domain $\D(\delta)$ given by the elements $a\in\A$ such that the limit exists.}
 the \emph{derivations} on $\cA$. By \emph{suitable}, we mean that the evolutions preserves not only the algebraic structures on $\cA$, but also specific \emph{physical} conditions. 

Further, the tensor product functor allows to define the whole tensor algebra $\T^r_s(\A)$ along with symmetric and skewsymmetric elements.  The set $\T^1_1(\A)\simeq{\rm End}(\A)$ possesses a composition law which makes it into an associative algebra, i.e. the algebra of linear operators with respect to the vector  space structure of $\A$. The set $\DD(\A)\subset{\rm End}(\A)$ of derivations in  $\A$ is a module over the centre $Z(\A)$, and a Lie algebra with respect to the commutator structure in ${\rm End}(\A)$ given upon antisymmetrising the composition as 
$$
[A,A']\,=\,A\circ A'-A'\circ A
$$
on a suitable common domain for  $A,A'\in{\rm End}(\A)$ if $\A$ is infinite dimensional. 
Analogously, if  $\A$ is non commutative  the antisymmetrization of the associative product gives $\A$ a non trivial Lie algebra structure, with the commutator defined by 
\beq
\label{eq29}
[a,a']=aa'-a'a.
\eeq  
It is also true that 
\beq
\label{eq31}
[a\,a', b]\,=\,a\,[a',b]\,+\,[a,b]\,a'
\eeq
for any $a,a',b\,\in\,\A$, that is the map $a\,\mapsto\,[a,x]$ is a derivation in $\A$ for any $x\in\A$.

\begin{rema}
The relation  \eqref{eq31} is quite interesting, and somehow reversing the path we outlined it is  possible to prove \cite{bj2002} that 
if $\{~,~\}$ is a bilinear map in $\cA$ is any unital associative ring which coincides with the two sided ideal generated by the derived algebra $$\tilde{\cA}\,=\,{\rm span}\{\mathbb I\,\oplus\,[a,b]:\,a,b\in\cA\},$$ such that the Leibniz rule is satisfied on each argument, i.e.
$\{ab,c\}=a\{b,c\}+\{a,c\}b$ and $\{a,bc\}=\{a,b\}c+b\{a,c\}$, then
$$
\{a,b\}\,=\,k[a,b]
$$ 
with $k\in Z(\cA)$.  
\end{rema}

An element $\delta\in\DD(\A)$ (the set of derivations on $\A$) is called \emph{inner} if there exists an element $x\in M_{\A}$ such that 
 \beq
 \label{eq24}
 \delta(a)\,=\,[a,x];
 \eeq
by $M_{\A}$ we denote the algebra of multipliers for $\A$, that is the set 
$$
M_{\A}\,=\,\{x\,\in\,\A^{**}\,:\,x\,a\quad{\rm and }\quad a\,x\,\in\,\A\quad\forall\,a\,\in\,\A\}
$$
with $\A^{**}=(\A^*)^*$ the double dual to $\A$. Derivations which are not inner are called \emph{outer}. 

We define the time evolution of a physical system associated to an algebra $\A$ to be \emph{quantum}  if the dynamics is given by  a one parameter group $U_t$ of  automorphisms for $\A$ whose infinitesimal generator is an inner derivation $\delta$ corresponding to a real element $H\in M_{\A}$ (the Hamiltonian of the system) so that we write the time evolution as
\beq
\label{eq25}
\frac{\dd a}{\dd t}\,=\,\dot a\,=\,-i[a,H]
\eeq
for $a\in\A$. We quote \cite{sudarshan-sdt1962}, 

\bigskip

{\it Our dynamical scheme is as follows: we have an abstract Lie algebra $\cL$ whose elements constitute the dynamical variables and a concrete linear associative algebra $\A$  which furnishes a realization of $\cL$ by derivations. The states (...) are normalised non negative linear functionals over $\A$.}

\bigskip

\subsection{Quantum and classical dynamics}\label{ssec2.3}In order to show how both quantum and classical dynamics fall into such a scheme, we recall (see \cite{GfD, klaas-fqt, gianfa85}) that in quantum mechanics one considers the observables $\cO$ as the (real linear space) of self-adjoint operators on a separable Hilbert space $\cH$, and each probability measure corresponding to a state $\rho\in\cS$ can be written as
$$
\mu(a,\rho)\,:\,\Delta\quad\mapsto\quad{\rm Tr}(W_{\rho}\,E_a(\Delta))
$$
where
$E_a(\Delta)$ is the (orthogonal) projection operator on $\cH$ given by the spectral resolution of $a$, for the subset of the spectrum $\sigma_a\in\Delta$, 
$W_{\rho}$ is a density operator on $\cH$, i.e. a positive operator with ${\rm Tr}\,(W_{\rho})=1$. 
 Pure states are identified with rank 1 density operators, i.e. rank 1 projection $$W_{\rho}^2=W_{\rho}=W_{\rho}^{\dagger}$$ operators on $\cH$; equivalently pure states can be identified with elements in the projective space $\mathbb P(\cH)\,=\,\cH_0/\C_0$. 
No state which is  dispersion free with respect to all observables exists. 

We analogously recall that in classical mechanics one starts upon identifying the observables as a  set of suitably regular ($\R$-valued) functions $\cO=\mathcal F(M)=\cA$ on a (smooth) manifold $M$, states $\rho\in\cS$ as the set of probability measures on $M$. 
Density states are given by absolutely continuous measures $\dd\mu$ (w.r.t. to a reference measure  $\dd m$ on $M$ for example induced by a Lebesgue measure on a Euclidean vector space $\R^{\dim M}$)   so that the pairing  is
$$
\mu(f,\rho)\,:\,\Delta\qquad\mapsto\qquad\int_{f^{-1}(\Delta)}\dd\mu_{\rho},
$$
pure states are given by singular Dirac's measures $\delta_m$, with pairing 
$$
\mu(f,\delta_m)\,:\,\Delta\qquad\mapsto\qquad\left\{\begin{array}{l} 1\,\quad{\rm if}\quad m\in\Delta \\ 0\quad{\rm if}\quad m\notin\,\Delta\end{array}\right. $$
Given a state $\rho$, the mean value and variance for any observable are
$$
\langle f\rangle_{\rho}\,=\,\int\dd\mu_{\rho}f\qquad\qquad \sigma^2_{\rho}(f)\,=\,\langle (f-\rho(f))^2\rangle_{\rho}.
$$
Pure states give dispersion free states for any observable. The duality we have mentioned above can be more properly analysed. 
The Riesz representation theorem proves that, given $\cA=C_0(M)$, then any positive continuous linear functional $\phi\in\cA^*$ can be represented as a regular measure on $M$. Equipped with the weak $^*$-topology, this set is compact and convex. Its extremals are the singular $\delta$-measures, i.e. the pure states. 
Given the algebra $\cA=\cB(\cH)$, the set of positive, bounded linear functionals $\phi\in\cA^*$ with unit norm can be written as $$\phi(A)={\rm Tr}(WA)$$ for a unique density operator $W=W^{\dagger}, \quad W\geq0,\quad {\rm Tr}\,W=1$ if and only if  it is normal.
The Gleason's theorem proves that well defined probability densities for quantum mechanics are still described in terms of density operators. Such a set is compact and convex, its extremals give the pure states, which can be identified with rank 1 projection on $\cH$. 

The G.N.S. theorem for $C^*$-algebras suggests a possible unified  setting for the formalisms introduced above \cite{klaas-fqt}. As well known, a $C^*$-algebra is a normed $*$-algebra $\A$ such that $\A$ is a Banach space with respect to the topology induced by the norm and such that $\|aa^*\|=\|a\|^2$ for any $a\in\A$. If $\A$ is a unital commutative $C^*$-algebra then there exists a compact Hausdorff topological space $M$ such that $\A$ is isometrically $*$-isomorphic to the $C^*$-algebra $ C(M)$ (with the uniform convergence norm); if $\A$ is a non unital commutative $C^*$-algebra, then $\A$ turns out to be isometrically $*$-isomorphic to $C_0(M)$, i.e. the algebra of continuous functions vanishing at infinity on a locally compact Hausdorff space $M$.  If $\A$ is a non commutative $C^*$-algebra, then there exists a separable Hilbert space $\cH$ such that $\A$ is isometrically $*$-isomorphic to a subalgebra  $\cA'\subseteq\cB(\cH)$ equipped with the usual operator norm.  

In order to analyse such a formalism, we begin by assuming  that the physical system we consider is associated to the $C^*$-algebra $\A\subseteq\cB(\cH)$ with $\cH=\C^n$ for $n<\infty$. If $\A$ is commutative, then one easily sees that there are no inner derivations (nor outer, indeed) since the automorphisms group ${\rm Aut}(\A)$ is finite. This is related to the fact that, upon a unitary conjugation, all elements in $\A$ can be simultaneuosly diagonalised, so that $\A\simeq\C^k$ with $k\leq n$, which is the algebra of functions on a $k$-points space.  
If $\A\subseteq\cB(\C^n)$ is maximally non commutative, then we say that it represent a \emph{finite level quantum system}. Since it can be proven that  the action of any one parameter group of automorphisms for $\cB(\cH)$ can be represented as a conjugation, i.e.
\beq
\label{eq26}
U_t(x)\,=\,u^*_t\,x\,u_t
\eeq
with $u_t$ a one parameter group of unitary elements in $\cB(\cH)$, any quantum dynamics on $\A$ can be written as
\beq
\label{eq27}
\dot a\,=\,-i[a,H]
\eeq
for any $a\in\A$ with $H\in M_{\A}\subseteq\cB(\cH)$ giving the Hamiltonian $H=H^{*}$ such that $u_t=e^{-itH}$ in terms of the usual exponential of bounded operators (finite dimensional matrices, in this case). One recovers in this way that \eqref{eq27} gives the quantum evolution within the Heisenberg picture, while dually
\beq
i\dot\psi\,=\,H\psi
\label{eq28}
\eeq
with $\psi\in\cH$ gives the corresponding Schr\"odinger equation. 

Although (see for example \cite{bra-book, bra-der})  the relation \eqref{eq26} is valid also for $\cB(\cH)$ for an infinite dimensional Hilbert space $\cH$, it is important  to notice that many observables in quantum systems are described by unbounded operators on an infinite dimensional Hilbert space $\cH$: even the operators closing the  canonical commutation relations, i.e. 
\beq
\label{eq30}
[Q_a,P_b]\,=\,i\delta_{ab}
\eeq
 can be defined  only on infinite dimensional Hilbert spaces and can not be both  bounded. The $C^*$-algebra formalism, as we discussed it above, seems to have no room to accomodate them. Nonetheless, if a physical system is associated to a set  $\tilde\A$ of (possibly unbounded) operators on a Hilbert space $\cH$ (that is we write $\tilde\cA\subseteq{\rm Op}(\cH)$), we say it has a quantum dynamics if the evolution is described, recalling the Wigner's theorem, by a one parameter group $u_t$ of unitary operators on $\cH$ so that the time evolution of any observable $a\in\tilde\cA$  is given by
\beq
\label{eq33}
a(t)\,=\,u^*_tau_t:  
\eeq
notice that, if $a$ is unbounded, this relation does not alter its domain. The infinitesimal version of such relation is clearly the Heisenberg equation, which can be  written on a dense domain as an inner derivation for $\tilde\A$, formally analogue to \eqref{eq27}
\beq
\label{eq34}
\dot a\,=\,-i[a,H]
\eeq
with $u_t=e^{-itH}$ for $H^*=H$.

Analogously, many observables in classical systems are described by functions which are not elements in $C_0(M)$ (for a non compact manifold $M$): consider for example  the velocity, or the momentum, of a point particle in a purely classical (i.e. not relativistic) setting.  
  
A possible path to describe in a general algebraic setting the \emph{canonical} formulation of both classical mechanics and quantum mechanics for a physical system is based  again on  Dirac's ideas. As we have already noticed, if  $\A$ is a (normed) non commutative $*$-algebra, the commutator defined as in \eqref{eq29} defines a Lie algebra structure and provides a set of inner derivations, via the identity \eqref{eq31}. Can an analogue structure be defined for the algebra $\cF(M)=C^{\infty}(M)$, which gives all the relevant observables for a classical system? Such a  structure on $\cF(M)$ should be a  map 
$$
\{~,~\}\,:\,\cF(M)\,\times\,\cF(M)\,\to\,\cF(M)
$$
such that 
\begin{align}
& \{f,f'\}=-\{f',f\}, \nn \\
& \{f+f',g\}\,=\,\{f,g\}\,+\,\{f',g\}, \nn \\
& \{f,\{g,h\}\}+\{g,\{h,f\}\}+\{h,\{f,g\}\}\,=\,0 \nn \\
& \{f\,f',g\}\,=\,f\{f',g\}+\{f,g\}f'
\label{eq36}
 \end{align}
for any $f,f',g,h\,\in\,\cF(M)$. Well, when this map exists, it is called\footnote{see \cite{izu}.} a Poisson structure on $M$. 
We limit ourselves to recall that if $M=T^*Q$, i.e. the manifold $M$ is the cotangent bundle (the phase space) of a configuration manifold $Q$, then there exists a canonical Poisson structure given locally, by 
\beq
\{q^a,p^b\}\,=\,\delta^{ab}
\label{eq32}
\eeq
while, for a general manifold $M$ with respect to a local system of coordinates $\{x^a\}_{a=1,\dots,\dim M}$, the Poisson tensor turns to be a bivector field  characterised by 
$$
\{x^a,x^b\}\,=\,\Lambda^{ab}
$$
with $$\Lambda^{ck}\del_k\Lambda^{ab}+\Lambda^{ak}\del_k\Lambda^{bc}+\Lambda^{bk}\del_k\Lambda^{ca}=0$$ (the Jacobi identity) and $\Lambda^{ab}=-\Lambda^{ba}$, so to have
\beq
\label{eq33bis}
\{f,f'\}\,=\,\Lambda^{ab}\frac{\del f}{\del x^a}\frac{\del f'}{\del x^b}.
\eeq
One can prove that ${\rm Aut}(\cF(M))$ is equivalent to the set of ${\rm Diff}(M)$, that is for any $\Phi\in{\rm Aut}(\cF(M))$ there exists a $\phi\in{\rm Diff}(M)$ such that 
$$
\Phi(f)\,=\,\phi^*(f)\,=\,f\circ\phi
$$
in terms of the pull-back functor. 
A theorem by Willmore proves that the set of derivations for the algebra $\cF(M)$, with respect to the local commutative pointwise product, coincides with the set $\mathfrak{X}(M)$ of vector fields on $M$. A derivation on $M$  can be  locally written as 
\beq
\label{eq35}
\mathfrak{X}(M)\,\ni\,\delta\,=\,\delta^a(x)\frac{\del}{\del x^a};
\eeq
being the  product in $\A=\cF(M)$ commutative and local, inner derivations do not exist. The analogy between the algebraic structure of the commutator (as in \eqref{eq29}-\eqref{eq31}) with the Poisson structure (as in \eqref{eq36}), analogy that we write as
$$
-i\,[\quad,\quad]\,\sim\,\{\quad,\quad\},
$$
 suggests that the role of inner derivations for a classical system can be played by Hamiltonian derivations, i.e. those vector fields on $M$ such that a function $H\in\cF(M)$ exists such that 
\beq
\label{eq37}
\delta(f)\,=\,\{f,H\}.
\eeq
We then say that the time evolution of a classical system  associated to $\cF(M)$ is Hamiltonian if the dynamics is described by a one parameter group of diffeomorphisms on $M$ whose infinitesimal generator is a Hamiltonian vector field, i.e. such that the differential equation describing the infinitesimal motion can be written as 
\beq
\label{eq38}
\frac{\dd f}{\dd t}\,=\,\dot f\,=\,\{f,H\}
\eeq
for suitable Hamiltonian function $H\in\cF(M)$. 
Under this respect, we notice that the Poisson formalism for a classical system naturally stands among the other well known descriptions (say the symplectic, the  Lagrangian, or the Newtonian ones) because of its close algebraic relations with the quantum formalism. 

\subsection{Classical and quantum integrability}\label{ssec2.4} This algebraic setting allows to analyse both  classical concepts within quantum mechanics, and quantum concepts within classical mechanics. 

Consider for example a physical system associated to the commutative algebra $\A=C^{\infty}(\R^{2N})$, whose dynamics is the free evolution given by 
\beq
\label{eq39}
(\Phi_t(f))(q,p)\,=\,f(q+tp, p)\,=\,\sum_{k=0}^{\infty}\frac{(tp)^k}{k!}\,\del_q^kf\,=\,(e^{t\delta}f)(q,p)
\eeq
with $f\in\A$ and $q,p$ collectively denoting the global coordinate system $(q^a,p^a)_{a=1,\dots,N}$: here 
\beq
\label{eq39b}
\delta=p^a\del_{q^a}
\eeq
 is the vector field (i.e. the derivation) generating the dynamics. If we analyse this example in algebraic terms, we see that $\A$ can be given as a suitable norm completion of the polynomial algebra on a finite set of generators, that is we can write $\A\,\simeq\,\C[q^a,p^a]_{a=1,\dots,N}$ and the derivation $\delta$ in \eqref{eq39b} is  
order two \emph{nihilpotent} when its action is restricted to the generators, with 
\begin{align}
&q^a\quad\stackrel{\delta}{\mapsto}\quad p^a,\nn \\ & p^a\quad\stackrel{\delta}{\mapsto}\quad 0. 
\label{eq39t}
\end{align}
This means that the exponential series development \eqref{eq39} upon generators truncates at a finite order. A more general  example in classical mechanics is given by  dynamics which is completely integrable. 
Consider the smooth manifold $M\simeq T^N\times \R^N$ (with $T^N$ the $N$-dimensional torus and $\R^N$ the quotient of $M$ under a $T^N$ foliating action \cite{mssv85}) and the evolution locally given by 
\beq
\label{eq40}
(\Phi'_tf)(\theta, I)\,=\,f(\theta+tI, I)\,=\,(e^{t\delta'}f)(\theta, I)
\eeq
with $f$ a function defined on a local chart $V\subset M$ where the  $\theta$ (i.e. the periodic \emph{angle})  variables are well defined   and $I$ denote the $\R^N$ (i.e. the \emph{action}) variables. We can then write 
\beq
\label{eq40b}
\delta'=I^a\del_{\theta^a}
\eeq
on $V$ and 
\begin{align}
&\theta^a\quad\stackrel{\delta'}{\mapsto}\quad I^a, \nn \\ & I^a\quad\stackrel{\delta'}{\mapsto}\quad 0. \label{eq41}
\end{align}
In order to give an algebraic description of this dynamics we recall  the notion of group valued functions as in \cite{dirac-nodes} and define\footnote{If $I$ denote an action variable on $M$ (which for simplicity assume as $M=\R\times {\rm S}^1$) an angle variable $\theta$ can be algebraically described in terms of a ${\rm U}(1)$ valued functions $u$ such that $\{u, I\}=iu$. If $a$ is a function on $M$ such that $\{I,\{I,a\}\}=-a$, then one can define $u=(a+ib)\{(a-ib)(a+ib)\}^{1/2}$ with $\{I,a\}=b$. 

Alternatively, one can think of the $u$ elements as the commutative limit of the algebra of the non commutative torus. }  the algebra $\cA\,\simeq\,\C[u^a=e^{i\theta^a},I^a]_{a=1,\dots,N}$. From the local \eqref{eq41}, the global expressions are
\beq
\label{eq41b}
e^{i\theta^a}\quad\stackrel{\delta'}{\mapsto}\quad (iI^a)e^{i\theta^a}, 
\eeq
so to have
\beq
\label{eq41t}
e^{i\theta^a}\quad\stackrel{\Phi'_t}{\mapsto}\quad e^{i(tI^a+\theta^a)}.
\eeq

 One can generalise this analysis along two different lines. If a classical system is associated to a commutative algebra $\A\simeq\C[x^a]_{a=1,\dots,N}$ and the dynamics is generated by a derivation $\delta\in\,\DD(\A)$ which is nihilpotent of order $k\geq1$ on generators\footnote{We mean that there exists a $k\,\in\,\mathbb{N}$ such that $\delta^k(x^a)=0$ for any generator of $\A$.}, then the exponential series truncates at a finite order upon generators, and we can say this dynamics represent an \emph{order $k$ integrable evolution}\footnote{We refer the reader to \cite{IMRT2019}, where a more complete analysis of nihilpotent integrability and associated reduction is developed within an algebraic setting, based on the notion of differentiable algebra \cite{ng-diff}. }. Moreover, this definition can be used also within the quantum formalism. If a physical system is associated with a non commutative algebra  given as a norm completion of a polynomial algebra $\A\simeq\C[x^a]_{a=1,\dots,N}$ where the generators no longer commute, and for  the derivation $\delta$ generating the dynamics there exists an order of nihilpotency,  then we can think of such evolution as an example of a \emph{quantum integrable} dynamics.

\subsection{About the Wigner's problem}\label{ssec2.5}Other interesting considerations arise. 
Since the equations of motions are written in terms of derivations on an algebra $\cA$, Wigner's problem \cite{wig50} appears natural: \emph{Do the equations of motions determine the quantum mechanical commutation relations?}
Given 
\begin{align*}
&\dot q=p, \\ & \dot p=-\del V/\del x
\end{align*} for elements in $\A\subseteq{\rm Op}(\cH)$, do we necessarily get \eqref{eq30}
$$[q,p]=i?$$ 
The answer is in the negative for both the free particle and the harmonic oscillator dynamics. 

Within the classical realm, the analogous question arises. If the dynamics is given by a vector field $\delta\in\mathfrak X(M)$, does a Poisson tensor $\Lambda$ exist, such that the relation \eqref{eq38} is satisfied, with 
$$
\delta(f)\,=\,\{f,H\}\,=\,\Lambda(\dd f, \dd H)
$$  
for a suitable Hamiltonian? This is a formulation of the so called inverse problem in the Poisson formalism. Assuming that a vector field is Hamiltonian with respect to a given Poisson structure on a manifold, one can generalise the problem posed by Wigner and wonder  \emph{do the equations  of the  motion determine the Poisson tensor?} Even this answer is in the negative\footnote{Notice also that the Euler derivation $\delta=x^a\del_{x^a}$ on $\R^N$ is not Hamiltonian with respect to \emph{any} Poisson structure on the whole $\R^N$. The antisymmetry of the Poisson structure implies that $\delta(H)=\{H,H\}=0$, so possible Hamiltonians for $\delta$ must be annihilated by $\delta$. It is indeed clear that the only continuous functions annihilated by $\delta$ are the constants.}. There are examples \cite{lpbv97} of a given vector field $\delta$ which is Hamiltonian w.r.t. different Poisson tensors $\Lambda, \Lambda'$, mapped one into the other under  a non canonical and non linear transformation on the carrier space $M$. 

We further consider the following situation. Consider again an algebra $\A\simeq\C[x^a]_{1,\dots,N}$ and a degree zero homogeneous derivation 
\beq
\label{eq43}
\delta\,:\,x^a\quad\mapsto\,\quad c^a_bx^b
\eeq
(with $c^a_b\in\C$). Such a derivation is extended to $\A$ by requiring it to be linear and to satisfy the Leibniz rule. Notice that, among the dynamics associated with this derivation, one has the free (as seen above in \eqref{eq41}) and the harmonic oscillator dynamics, given by
\begin{align}
&q^a\quad\stackrel{\delta}{\mapsto}\quad p^a,\nn \\ & p^a\quad\stackrel{\delta}{\mapsto}\quad -\omega^2q^a.
\label{eq42}
\end{align}
We observe that the equations of the motions \eqref{eq43} \emph{do not even determine whether the associative product in $\A$ is commutative or not.} 

In order to further qualify what we say, we turn our attention to the 
the Weyl-Wigner formalism which, together with the notion of  tomograms  (see \cite{tomo-team}) and  that of  deformation quantization (see \cite{def-team, cahen-gutt})   allow to describe a quantum system  in terms of an algebra of functions on a classical manifold $M$ and  the transition from the $C^*$-algebra $C_0(M)$ to a suitable subalgebra of smooth elements in  $\cF(M)$. 

We briefly sketch it for the case of $M=\R^{2N}$ equipped with the canonical Poisson tensor $$\{q^a, p^b\}=\delta^{ab}$$ as in \eqref{eq32} along a global Darboux coordinate system given by $(q^a, p^a)_{a=1,\ldots,N}$. Such a Poisson tensor is non degenerate, and one can define the corresponding\footnote{We limit ourselves here to recall that, if the Poisson tensor $\Lambda$ with local expression  $\Lambda=\Lambda^{ab}\del_{x_a}\wedge\del_{x_b}$ is non degenerate on $M$, then the corresponding symplectic structure $\omega$ can be defined as $\omega=\omega_{ab}\dd x^a\wedge\dd x^b$ with $\Lambda^{ab}\omega_{bc}=\delta^a_c$.} symplectic structure, given as the 2-form 
$$
\omega\,=\,\dd q^a\wedge\dd p^a.
$$
A Weyl system is a unitary projective representation $D\,:\,V\,\to\,\mathcal U(\mathcal H)$  of the abelian vector  group $(V, +)$ on a separable Hilbert space, such that
$$
D(v_1)D(v_2)D^{\dagger}(v_1)D^{\dagger}(v_2)\,=\,e^{i\omega(v_1,v_2)/\theta}.
$$
Via such a set  of so called Displacement operators one defines, on a suitable domain,  the map $W\,:\,\rm{Op}(\mathcal H)\,\to\,\mathcal F(\R^{2N})$ given (we denote by $\{z\}$ the coordinate functions  on the phase space $V=\R^{2N}$ and by $\{w\}$ their Fourier dual coordinates) as
$$
W_A(z)\,=\,\int_{\R^{2N}}\frac{\dd w}{(2\pi\hbar)^N}\, e^{-i\omega(w,z)/\theta}\,{\rm Tr}[A\,D^{\dagger}(w)]
$$
that associates, to a suitable operator $A$ on $\mathcal H$, its \emph{Wigner symbol}, i.e. a function $W_A$ on the classical phase space $\R^{2N}$. With $W$  proven to be injective, the non commutative Moyal algebra is defined as the set of  Wigner symbols equipped with the product  given  by  $$(W_A*W_B)(z)=W_{AB}(z).$$ 
It reads (see \cite{marse, alg1, alg2})  (with $\theta>0$)  
\beq
\label{s1}
(f*g)(x)\,=\,\frac{1}{(\pi\theta)^{2N}}\int\int\dd u\dd v \,f(x+u)g(x+v)\,e^{-2i\omega(u,v)/\theta}
\eeq
 for $f,g\,\in\,{\mathcal S}(\R^{2N})$, i.e. the Schwartz space in $\R^{2N}$. This means that, on the algebra 
 ${\mathcal S}(\R^{2N})$ we have \emph{both} the commutative local pointwise product $fg(x)=f(x)g(x)$ \emph{and} the Moyal product.
 Such a product is non local and non commutative.
 The set 
$\mathcal A_{\theta}\,=\,(\mathcal S(\R^{2N}), *)$
is a non unital pre - $C^*$-algebra. Via the tracial property of this algebra, it is possible to define the  Moyal product of a tempered distribution $\mathcal S'(\R^{2N})$ times a test function and then to 
consider the space of left and right multipliers
\begin{align}
&\mathcal M_{L}^{\theta}\,=\,\{T\,\in\,\mathcal S'(\R^{2N})\,:\,T*f\,\in\,\mathcal S(\R^{2N})\,\forall \,f\,\in\,\mathcal S(\R^{2N})\} \nn \\
&\mathcal M_{R}^{\theta}\,=\,\{T\,\in\,\mathcal S'(\R^{2N})\,:\,f*T\,\in\,\mathcal S(\R^{2N})\,\forall \,f\,\in\,\mathcal S(\R^{2N})\}; 
\label{s3}
\end{align}
the set $\mathcal M^{\theta}=\mathcal M_{L}^{\theta}\cap\mathcal M_{R}^{\theta}$ is a unital $*$-algebra that contains polynomials, plane waves, Dirac's $\delta$ and its derivatives. Its classical limit 
\beq
\lim_{\theta\to 0}\,\mathcal M^{\theta}\,=\,\mathcal O_{M}
\label{s4}
\eeq
is the set of smooth functions of polynomial growth on $\R^{2N}$ in all derivatives. The Moyal product has, on a suitable subset of $\mathcal M^{\theta}$, the asymptotic expansion in $\theta$ given by
\beq
\label{pros1}
f*g\,\sim\,fg\,+\,\frac{i\theta}{2}\{f,g\}\,+\,\sum_{k=2}^{\infty}(\frac{i\theta}{2})^k\frac{1}{k!}D_{k}(f,g)\qquad\mathrm{as}\,\,\theta\,\,\to\,\,0
\eeq
with $D_k$ the $k$-th order bidifferential operator which  is written as\footnote{For the more general problem of defining a deformed product on a general Poisson manifold, we refer to \cite{maxim}.}
\beq
\label{trao}
D_{k}(f,g)\,=\,\frac{\del^kf}{\del q^k}\,\frac{\del^kg}{\del p^k}\,-\,\left(\begin{array}{c} k \\ 1\end{array}\right)\,\frac{\del^kf}{\del^{k-1}q\del p}\,\frac{\del^k g}{\del^{k-1}p\del q}\,+\,\dots\,+\,(-1)^k\,\frac{\del^k f}{\del p^k}\,\frac{\del^kg}{\del q^k}. 
\eeq
If $f,g\in\mathcal M^{\theta}$,  then 
\beq
\label{mobra}
[f,g]\,=\,f*g-g*f\,=\,i\theta\,\{f,g\}\,+\,\sum_{s=1}^{\infty}\frac{2}{(2s+1)!}\,\left(\frac{i\theta}{2}\right)^{2s+1}D_{2s+1}(f,g):
 \eeq
we see that the Moyal (i.e. quantum) product is a deformation of the pointwise standard product, and that the (quantum)  commutator is a deformation of the Poisson bracket. We can then assume $\mathcal O_M$ as the algebra of classical observables, and $\mathcal M^{\theta}$ as the algebra of quantum observables: the classical evolution can be seen as a limit ($\theta\to0$, with $\theta$ being  $\hbar$ in disguise) of the quantum evolution within such an algebraic formalism. 

One proves that the Moyal space is a so called normal space of distribution, i.e. all derivations are inner. In the commutative limit,  Moyal derivations are mapped into Hamiltonian vector fields.  
For degree 1 polynomials we have the canonical commutation relations 
$$
[q^a,q^b]_{\theta}=0, \qquad [p^a,p^b]_{\theta}=0, \qquad  [q^a, p^b]_{\theta}=i\theta\delta^{ab}
$$
while, if $f,g\,\in\,S=\mathcal P_0\oplus\mathcal P_1\oplus\mathcal P_2$ (with $P_k$ the vector space of degree $k$ homogeneous polynomials on $\R^4$)
$$
[f,g]\,=\,i\theta\{f,g\}.
$$
We focus now on $M=\R^4$. It is immediate to see that the vector space  $(S, \,\{~,~\})$ is a Poisson subalgebra of $\mathcal F(\R^4)$, while $(S, [~,~]_{\theta})$ is a Lie subalgebra in  $\mathcal M^{\theta}$ w.r.t. the $*$-product commutator. Such a space is   isomorphic to a one dimensional central extension of  the Lie algebra $\mathfrak{isp}(4,\R)$  corresponding to the inhomogeneous  symplectic linear  group, and we notice that $$(S, [~,~]_{\theta})\sim(S,\{~,~\})$$ is the maximal Lie algebra acting upon both $\mathcal F(\R^4)$ and $\mathcal M^{\theta}$ in terms of derivations. 
This means that, if we have an operator $\delta$ in $\cF(\R^4)$ whose action is \eqref{eq43} $$\delta\,:\,x^a\mapsto c^a_bx^b,$$ then we can extend it \emph{both} as a derivation in $\cM^{\theta}$ and as a derivation in $\cO_M$.
We can (between quotes) then say that there exist derivations on $\cF(\R^4)$ which do not  determine whether the product in the algebra of observables is commutative or not.

\section{Symmetries and reduction for a classical dynamics}
\label{sec3}

Consider a classical physical system described by a commutative algebra $\A\subseteq\cF(M)=C^{\infty}(M)$ with a dynamics given by the derivation $\delta\in\DD(\A)\subseteq\mathfrak X(M)$. A description for the reduction of such a dynamics is as follows (see \cite{GfD, glmv94, noether1})

Let $$F\,:\,M\,\to\,M'$$ be a map between the manifold $M$ and a $N'$-dimensional smooth manifold $M'$ with $N>N'$. If a vector field $\delta_{(F)}$ on $M'$ exists, such that it is $F$-related to $\delta$, i.e. 
\beq
\label{eq2}
F_*(\delta)=\delta_{(F)},
\eeq 
then we say that $F$ \emph{reduces} $\delta$. If such a map $F$ exists for a given dynamics $\delta$, then we can define the set 
 \beq
 \label{eq3}
 \mathcal A_F\,=\,\{F^*(f')\,:\, f'\in\mathcal F(M')\}\,\subset\,\A,
 \eeq
which turns to be a  subalgebra in $\A$   invariant under the dynamics, namely $\delta\in\DD(\A_F)$. 
The time evolution of the elements in $\mathcal A_F$ the time evolution generated by $\delta$ on the whole $\A$, supposedly easier to solve. 
Given the map $F$ we can also define  
\beq
\label{eq4}
\mathcal D_F\,=\,\ker\,F_*\,=\,\{Y\,\in\,\mathfrak X(M)\,:\,Y(f)=0\quad\forall \,f\,\in\,\mathcal A_F\}, 
\eeq
which is an infinite dimensional Lie subalgebra in $\mathfrak X(M)$.  It means that $\mathcal D_F$ is an involutive, i.e. an integrable distribution, with  $[\delta, Y]\,\in\,\mathcal D_F$ for any $Y\,\in\,\mathcal D_F$. The integral manifolds of the distribution $\mathcal D_F$ can be identified with the level sets of the map $F$, namely 
\beq
\label{eq8}
N_{m'}=\{m\in M:F(m)=m'\in M'\}.
\eeq
The quotient space given by identifying points on $M$ belonging to the same $N_{m'}$  is locally homeomorphic to $M'$. The flow generated by $\delta$ maps leaves $N_{m'}$ into leaves, the dynamics along each leaf is to be determined. 

Determining, for a given dynamics $\delta$ on $M$,  a suitable manifold $M'$ and a map $F\,:\,M\,\to\,M'$ that reduces it is highly non trivial. 
A possible strategy to solve it  consists in looking for a subalgebra (with respect to the associative pointwise product) ${\mathcal A'}\subset\mathcal A$ which is invariant under $\delta$, that is $\delta(f')\in{\mathcal A'}$ for any $\in\mathcal A'$. 
The set of derivations $X$ of $\A$  satisfying the condition $$X(f')=0$$ for any $f'\in\cA'$  thus providing an involutive distribution $\mathcal D_{\mathcal A'}$.  
When the distribution $\mathcal D_{\mathcal A'}$ is regular, it gives a regular foliation $\Phi_{\cA'}$ whose leaves can be identified with submanifolds  of constant (say $N-k$) dimension, and the quotient $M/\Phi_{\cA'}\simeq M'$ as a manifold structure:  the map $F$ is then recovered as the submersion associated to the foliation $\Phi_{\cA'}$ and ${\mathcal A'}\simeq\mathcal F(M/\Phi_{\cA'})$.


Aiming at a  generalisation of such analysis  to the problem of a reduction of a quantum dynamics, we describe it in terms of infinitesimal generators. 
Given an involutive distribution  $\mathcal D$ on  $M$ (i.e. $[Y,Y']\in \mathcal D$ for any pair $Y,Y'\in\mathcal D$) one can define its \emph{normaliser} 
\beq
\label{eq7}
\mathrm N_{\mathcal D}=\{X\in\mathfrak X(M)\,:\,[X,Y]\in\mathcal D\quad\forall \,Y\,\in\, \mathcal D\},
\eeq  so to have the short exact sequence of Lie modules
\beq
\label{eq5}
0\quad\rightarrow\quad\mathcal D\quad\rightarrow\quad {\rm N}_{\mathcal D}\quad\rightarrow\quad{\rm R}_{\mathcal D}\quad\rightarrow\quad 0:
\eeq
vector fields in ${\rm N}_{\mathcal D}$ can be reduced with respect to the distribution $\mathcal D$, the elements in ${\rm R}_{\mathcal D}$ 
give the equivalence classes of vector fields on $M$ whose projections onto the quotient manifold $M/\Phi_{\mathcal D}$ coincide. A dynamics $\delta$ can be reduced via the involutive distribution $\mathcal D$ if $\delta\in\rm N_{\mathcal D}$. In such a case, the set 
\beq
\label{eq6}
\mathcal A_{\mathcal D}=\{f\in\mathcal F(M)\,:\,Y(f)=0\,\,\forall\,\,Y\,\in\,\mathcal D\}
\eeq
is an invariant subalgebra for $\delta$, whose Gelfand spectrum defines $M'$ and clearly coincides with the space given by indentifying  elements in $M$ on the same integral submanifold of $\mathcal D$. When $\mathcal D$ is regular, $M'$ has a manifold structure and 
the map $F$ is recovered as the submersion associated to the foliation generated by $\mathcal D$. 

Assume that the regular Lie module $\D$ has rank $h<N$ and  is spanned by a set $\{Y_j\}_{j=1,\dots,h}$ of elements in $\mathfrak X(M)$, so that the the subalgebra $\A_{\D}$ \eqref{eq6} can be written as
\beq
\label{eq9}
\A_{\D}\,=\,\{f\,\in\,\F(M)\,:\,Y_j(f)\,=\,0\}
\eeq
When the set of 1-forms $f_a\dd f'_a$ with $f_a,f_a'\,\in\,\A_{\D}$ generate at each point in $M$ a subspace with dimension  $N-h$, one can decompose $M$ into a set of smoothly parametrised submanifolds of dimension $h$.  Each leaf of the foliation $\Phi_{\D}$ generated by $\D$ is locally diffeomorphic to  a $h$-dimensional manifold $H$, while the elements in $M'$ parametrise the quotient $M/\Phi_{\D}$, and we can locally write
\beq
\label{eq10}
M\,\simeq\,H\times M'.
\eeq
Select a set of 1-forms $\{\alpha^j\}_{j=1,\dots,h}\,\in\,\Lambda^1(M)$ such that\footnote{Recall that an exterior form $\alpha\in\Lambda(M)$ is \emph{horizontal} with respect to the fibration infinitesimally generated by  $\D$ if $i_Y\alpha=0$ for any $Y\in\D$ and one can identify the exterior algebra $\Lambda(M')\simeq\{\alpha\in\Lambda(M)\,:\,i_Y\alpha=0,\,i_Y\dd\alpha=0\,\,\forall\,Y\,\in\,\D\}$. Such forms on $M$ are also called \emph{basic}. }
\beq
\label{eq11}
i_{Y_j}\alpha^k\,=\,\delta^k_j
\eeq
and define the $(1,1)$-tensor field 
\beq
\label{eq11b}
P\,=\,Y_j\otimes\alpha^j,
\eeq
which acts on $X\in\mathfrak X(M)$ as $$P(X)\,=\,(i_X\alpha^j)Y_j.$$ It is immediate to see that $P\circ P=P$, and the range of the action of $P$ upon $\mathfrak X(M)$ is the distribution $\D$, which coincides with $\ker(1-P)$. We say $P$ is a \emph{generalised connection} on $M$, with $$P(Y_j)=Y_j.$$ Notice that the choice of $P$ is not unique, since the 1-forms $\alpha^j$ satisfying the condition \eqref{eq11} are defined up to arbitrary horizontal 1-forms in $M$ with respect to  $\D$. Such a connection is called invariant along $\D$ if $L_{Y_j}P=0$. 

As we mentioned above, a classical dynamics $\delta$ can be reduced along $\D$, i.e. $\delta$ is compatible with the decomposition of $M$ associated to the regular involutive distribution $\D$ if $[\delta, Y]\,\in\,\D$ for any $Y\in\D$, equivalently if a set of functions $h_{j}^k\in\F(M)$ exists, such that $L_{\delta}Y_j=h_j^kY_k$. Under this condition, the subalgebra \eqref{eq6} $\A_\D\simeq\F(M')$ is invariant under the action of $\delta$, i.e. $L_\delta(f)\in\A_\D$ if $f\in\A_\D$.

A decomposition of a compatible $\delta$ induced by the regular involutive distribution $\D$ is given by 
\begin{align}
&\delta^\D=P(\delta), \label{eq12} \\
&\delta'=\delta-\delta^\D \label{eq13}
\end{align}
with $\delta'\,\in\,\DD(\cA_{\mathcal D})$, that is $\delta'(f)\in\cA_{\mathcal D}$ if $f'\in\cA_{\mathcal D}$. Once the evolution $\Phi'_t$ generated by $\delta'$ on $M'$ is determined, one has a reduced problem on $\mathcal D$ depending by $\Phi'_t$ as time dependent parameters. When it is 
possible to select $P$ in such a way that 
\beq
\label{eq14}
[\delta^\D,\delta']=0
\eeq
we may say the dynamics has been decomposed into independent motions, and we may solve separately the equations for $\delta^\D$ and the equations for $\delta'$. A particular case of this latter situation is given when $\delta'=0$. In this case we say that the decomposition is given in terms of generalised constant of the motion, thus recalling that $L_\delta f=0$ for any $f$ such that $L_{X_j}f=0$. When this last condition is satisfied, it is clear that dynamical vector field $\delta$ can be written as a combination of the $X_j$. This shows that there are plenty of evolutionary systems compatible with the given decomposition. 

Particular instances of the above procedure arise when, for example, a closed Lie group $H$ has a right (say) and proper action on a smooth manifold $M$, so that the quotient $M'\simeq M/H$ is the basis of a principal $H$-bundle with total space $M$. The vector fields $Y_j$ spanning the distribution $\D$ are the infinitesimal generators of the action of $H$ on $M$, and close a Lie algebra isomorphic to the Lie algebra $\mathfrak h$ of $H$. A natural example of this construction comes when $M$ is itself a Lie group manifold $G$ and $H$ is an embedded closed subgroup. The basis manifold $G/H$ is given by the so called lateral classes (left, or right). 

Peculiar examples of the described procedure emerge when the carrier manifold $M$  is the tangent or the cotangent bundle of a Lie group. In this case new structures appears in the general scheme and allow for additional analysis: the celebrated momentum map is made available and one may recover the so called Marsden-Weinstein reduction procedure.  Vector fields introduced above can be required to be Hamiltonian, as well as $\delta$ can be. When the carrier manifold is the tangent bundle of a Lie group, it is possible to consider dynamical vector fields which are Lagrangian, and again the reduction procedure will be coherent with such structures.

Our arguments find a quantum realisation when our Lie group is replaced by a unitary representation on a Hilbert space or as a group of automorphisms acting on a $C^*$-algebra $\A$.  In the classical case one can  consider the commutative $\A=\mathcal F(G)$, in the quantum case one can consider the $C^*$-algebra $\A$ defined as the group algebra on $G$ and its representations.

\section{A reduction scheme for  quantum dynamics}
\label{sec4}

In the previous section we described how it is possible to reduce a classical dynamics within an algebraic setting, and the conditions under which such reductions provides a decomposition of the dynamics generating an evolution which is given by the composition of independent motions. Along with the cartesian product decomposition \eqref{eq10}  the use of a suitable connection on the carrier manifold $M$ allows to write the decomposition
\beq
\label{eq44}
\cF(M)\,\simeq\,\cF(H\,\times\,M')\,\simeq\,\overline{\cF(H)\,\otimes\,\cF(M')}\,\supseteq\,\overline{\cF(H)}\,\otimes\,\overline{\cF(M')}
\eeq
with, if the condition \eqref{eq14} is satisfied, 
\beq
\label{eq45}
\delta\,=\,\delta^{\mathcal D}\otimes\,\mathbb{I} \,+\, \mathbb{I}\,\otimes\,\delta'
\eeq
The relations \eqref{eq44}-\eqref{eq45} show that, under the conditions we considered, a classical dynamical system described by the derivation $\delta$ acting on  the (commutative with respect to the pointwise product) algebra $\A\,\subseteq\,C^{\infty}(M)$ 
has been decomposed into lower dimensional elementary dynamical systems. Notice that such lower dimensional dynamics may be \emph{more} difficult to integrate: reducing a  linear dynamics via non linear constants of the motion may result in a non linear reduced dynamics, although on a lower dimensional carrier manifold (see \cite{GfD} for many examples). 

A possible scheme for the reduction of a finite level quantum dynamics written as the derivation 
\beq
\label{eq47}
\delta_H(A)\,=\,[A,H]
\eeq
 could start  as follows.
Consider a Lie subalgebra $\mathfrak g\subset\DD(\A)$, such that $[H,Y]\in\mathfrak g$ for any $Y\in\mathfrak g$. Define the set 
$$F_{\mathfrak g}=\{f\,\in\,\A\,:\,[f,Y]=0,\,\forall\,Y\,\in\,\mathfrak g\}.$$
 The set $F_{\mathfrak g}$ is both a subalgebra (with respect to the associative product) and a Lie subalgebra, playing the role of the algebra defined in \eqref{eq6} within the classical setting.

 The derivation $\delta_H$ (i.e. the dynamics) is a derivation for $F_{\mathfrak g}$ and also for $U_{\mathfrak g}$, which is the polynomial algebra generated by $\mathfrak g$.

 \begin{example}
 \label{exa1}
 An elementary example of this formalism arises when considering $\A=\cB(\C^N)$ with $U=U_{\mathfrak g}$ given by the block diagonal subalgebra 
\beq
\label{eq48}
U\,=\,\{u\,=\,\begin{pmatrix} \gamma & 0 \\ 0 & 0 \end{pmatrix}\,:\,\gamma\in{\rm Mat}_k(\C)\}
\eeq
with $k<N$. A derivation $\delta=\delta_H$ in $\cA$  maps elements in $U$ into elements in $U$ if and only if 
\beq
\label{eq49}
H\,=\,\begin{pmatrix} H_U & 0 \\ 0 & H_F\end{pmatrix},
\eeq
the algebra $F$ given by the commutant of $U$ is
\beq
\label{eq50}
F\,=\,\{f\,=\,\begin{pmatrix} \mathbb I & 0 \\ 0 & \phi \end{pmatrix} \,:\,\phi\in{\rm Mat}_{N-k}(\C)\}.
\eeq
The Hilbert space $\cH=\C^N\,=\,C^{N-k}\oplus\C^k$ has been decomposed into a direct sum, the dynamics does not mix the two subspaces. For what concerns the integration problem, we are now able to integrate low dimensional systems: in this sense we can say $\delta$ has been reduced, but the system has not been decomposed into more elementary systems.
\end{example}

\subsection{A differential calculus on a non commutative space}
\label{ssec4.2}
The general problem of analysing under which conditions a quantum dynamics can be decomposed into more elementary quantum dynamics (by \emph{decomposed into elementary} we mean that the algebra is decomposed into a tensor product of algebras with independent time evolutions) is beyond the scope we have in this paper. We limit ourselves to notice that the splitting we considered in the classical setting depends on the dynamics and therefore implicitly assumes the existence of a differentiable (smooth, indeed) structure on $M$. We turn now our attention to describe how it is possible to define a \emph{derivation based} differential calculus on an algebra $\A$ (see 
\cite{dv01,dica,dica2,segal67,segal70}).

The set $\underline{\Lambda}^k(\A)$ of $Z(\A)$-multilinear alternating maps (with $X_j\in\DD(\A)$)
$$
\omega\,:\,X_1\wedge\dots\wedge\dots X_k\,\mapsto\,\omega(X_1,\dots,X_k)\,\in\,\A
$$
 is the set of k-forms, with $\underline\Lambda^0(\A)\simeq\A$. 
 On the graded vector space $\underline{\Lambda}(\A)\,=\,\oplus_k\underline\Lambda^k(\A)$ one can define a wedge product by (with $\omega\in \underline{\Lambda}^j(\A)$ and $\omega'\in\underline{\Lambda}^{j'}(\A)$) 
\beq
\label{eq16}
(\omega\wedge\omega')(X_1,\dots,X_{j+j'})\,=\,\frac{1}{j!j'!}\,\sum_{\sigma\in \mathcal S_{j+j'}}(\mathrm{sign}(\sigma))\omega(X_{\sigma(1)},\dots,X_{\sigma(j)})\omega'(X_{\sigma(j+1)},\dots,X_{\sigma(j+j')})
\eeq
(where $\mathcal S_{j+j'}$ is the set of permutations of $j+j'$ elements). Each set $\underline{\Lambda}^k(\A)$ is then a $\A$-bimodule. The   operator $\dd:\underline{\Lambda}^n(\A)\to \underline{\Lambda}^{n+1}(\A)$ defined  by 
\begin{align}
(\dd\omega)(X_0, X_1,\dots, X_k)&=\,\sum_{k=0}^n(-1)^kX_k\left(\omega(X_0,\dots,\hat{X}_k,\dots,X_n)\right)\, \nn \\ &\qquad +\,\sum_{r<s}(-1)^{r+s}\omega([X_r,X_s], X_0,\dots, \hat{X}_r,\dots,\hat{X}_s,\dots, X_n)
\label{eq17}
\end{align}
(with $\hat{X}_r$ denoting that the $r$-th term is omitted)  is easily proven to be a graded antiderivation with $\dd^2=0$, so $(\underline{\Lambda}(\A), \dd)$ is a graded differential algebra.  Although these relations  are valid for both commutative and non commutative algebras, when the algebra $\mathcal A$ is not commutative one easily sees that it is  in general $A_1\dd A_2\neq (\dd A_2)A_1$ and $\omega\wedge\omega'\neq-\omega'\wedge\omega$. It is  indeed 
\begin{align}
\label{eq18}
&A_1\dd A_2\,:\,X\,\mapsto\,A_1(X(A_2))  \\ 
 \label{eq19}& (\dd A_2)A_1\,:\,X\,\mapsto\,(X(A_2))\,A_1,\end{align} 
  while 
   $$(\omega\wedge\omega')(X_1,X_2)\,=\,\omega(X_1)\omega'(X_2)-\omega(X_2)\omega'(X_1)$$ 
   and $$(\omega'\wedge\omega)(X_1,X_2)\,=\,\omega'(X_1)\omega(X_2)-\omega'(X_2)\omega(X_1).$$
This exterior algebra is an example of a derivation based calculus. Its subset $\Lambda(\A)$ is  defined as the smallest differential graded subalgebra of $\underline{\Lambda}(\A)$  generated in degree $0$ by $\mathcal A$. By construction, every element in $\Lambda^n(\A)$  can be written as a sum of $A_0\dd A_1\wedge\dots\wedge\dd A_n$ terms with $A_j\in\mathcal A$, while this is not necessary\footnote{One can indeed prove that, given the $\cA$-bimodule  $\Lambda(\cA)$, its dual module is the $\Z(\cA)$-module $\DD(\cA)$; analogously, the dual module to $\DD(\cA)$ turns out to be $\underline{\Lambda}(\cA)$. If $\cA\simeq\cF(M)$ with $M$ a paracompact manifold, then $\Lambda(\cA)\simeq\underline{\Lambda}(\cA)$.}    for elements in $\underline{\Lambda}(\A)$.

Upon the graded differential algebra $\underline{\Lambda}(\A)$ a contraction operator can be defined. If $X\in\DD(\A)$, then 
\beq
\label{eq20}
(i_{X}\omega)(X_1,\dots,X_n)\,=\,\omega(X,X_1,\dots,X_n)
\eeq
 gives a degree $(-1)$ antiderivation from $\underline{\Lambda}^{n+1}(\A)$ to $\underline{\Lambda}^n(\A)$.  The operator defined by $$L_{X}=i_X\dd+\dd i_X$$ is the degree zero Lie derivative along $X$, so we have a Cartan calculus on $\A$.

We begin by considering a finite level quantum system with $\cA=\cB(\cH)$ for $\cH\simeq\C^N$. We already know that all derivations for $\cA$ are inner, so if $\delta$ is a derivation for $\cA$, then there exists an element $X\in\cA$ such that $\delta(A)=[A,X]=\delta_X(A)$. Notice that we can write the relations \eqref{eq18}-\eqref{eq19} as 
\begin{align}
\label{eq18b}
&A_1\dd A_2\,:\,X\,\mapsto\,A_1\,[A_2,X] \\ \label{eq19b}& (\dd A_2)A_1\,:\,X\,\mapsto\,[A_2,X]\,A_1,\end{align}
and 
$$
L_XA\,=\,\dd A(X)\,=\,[A,X]\,=\,\delta_X(A). 
$$
This differential calculus contains all the differential calculi that can be defined upon considering Lie subalgebras $\mathfrak g\subset\A$. Moreover, let $\{X_j\}_{1,\dots,N^2}$ be a vector space basis for $\DD(\A)\simeq\A$: we can define the 1-forms $\{\alpha^j\}_{j=1,\dots,N^2}$ 
via 
\beq
\label{eq 21}
\alpha^j\,:\,X_k\,\mapsto\,\delta_k^j\mathbb I.
\eeq
A 1-form $\omega\in\underline{\Lambda}^1(\A)$  whose action is\footnote{Consider any 1-form $\omega\in\underline\Lambda^1(\A)$ such that $\omega(\mathbb{I})\neq0$. From $A_s\dd B_s(\mathbb{I})=A_s[\mathbb{I},B_s]=0$ it is evident that $\omega\not\in{\Lambda}^1(\A)$.}
$$
\omega\,:\,X_j\,\mapsto\,\omega^k_jX_k
$$
(with $\omega^k_j\in\C$) can be written as 
$$
\omega=\omega^k_jX_k\alpha^j,
$$
while, for any exact 1-form, it is
\beq
\label{eq22}
\dd A\,=\,[A,X_j]\alpha^j\,=\,(L_{X_j}A)\alpha^j
\eeq
Since the commutator gives traceless matrices, it is clear that  $\alpha^j$ is not exact, and 
\beq
\label{eq23}
\dd\alpha^j(X,X')\,=\,-\alpha^j([X,X'])
\eeq
gives an analogue of a Maurer-Cartan relation. 

Equipped the algebra $\A=\cB(\C^N)$ with such a differential calculus, we can consider from a more general point of view the example \ref{exa1} considered in the previous section.  We say that $\tilde P\,:\,\DD(\A)\,\to\,\DD(\A)$ is a generalised connection for the Lie algebra $U=U_{\mathfrak g}$ \eqref{eq48} if $\tilde P$ is a $Z(\cA)$-linear map such that $\tilde P^2=\tilde P$ with $\ker\,\tilde P=U$; we say it is invariant along $U_{\mathfrak g}$ if $[P(X), Y]=0$ for any $Y\,\in\,U_{\mathfrak g}$. Such a projection map allows to split the set $\DD(\A)$ into the direct sum of a vertical subspace (span by $U_{\mathfrak g}$) and a horizontal complement. This definition of a connection is equivalent to the one given in \eqref{eq11}-\eqref{eq11b} within the classical (commutative) setting, and generalises it to a setting where one has to deal with modules over non commutative algebras. it is immediate to see that the derivation in \eqref{eq49} can be written as the sum 
 \beq
 \label{eq51}
\delta_H\,=\,\delta_{H_U}+\delta_{H_F}
\eeq
where $\delta_{X_U}=(1-\tilde P)\delta_H$ and $\delta_{H_F}=\delta_H-\delta_{H_U}$.

To conclude, we limit ourselves to mention that a possible extension to the  infinite dimensional case comes  by considering the Moyal algebra. A differential calculus on $\mathcal M^{\theta}$ is defined by the inner
derivation operators
$$
\del_{q_a}f\,=\,-\frac{i}{\theta}[f,p_a], \qquad
\del_{p_a}f\,=\,\frac{i}{\theta}[f,q_a].
$$
They give a basis for the tangent space to $\mathcal M^{\theta}$.
For such vectors there exists a frame, i.e. a dual basis of 1-forms that we denote by $(\dd q_a, \dd p_a)_{a=1,2}$. They generate the whole exterior algebra over $\mathcal M^{\theta}$:
\begin{align*}
\dd f\,=&(\del_{q_a}f)*\dd q_a\,+\,(\del_{p_a}f)*\dd p_a \\
=&\frac{i}{\theta}\,([p_a, f]*\dd q_a\,+\,[q_a,f]*\dd p_a)
\end{align*}

Using the so called Jordan-Schwinger map, we can realise any 3d Lie algebra as a subalgebra in $(S,\,[~,~])$.
Any 3d Lie algebra $\mathfrak g$ gives a canonical Poisson bracket $\Lambda_{\g}$ on the dual $\tilde \g^*$. Each Poisson tensor $\Lambda_{\mathfrak g}$ has a Casimir 1-form $\alpha_{\g}$. We select those Lie algebras whose Casimir 1-form is exact, so that we have $\alpha_{\g}=\dd C_{\g}$.
For each of these Lie algebras, we define a non commutative algebra
$$
\tilde A_{\g}\,=\,\{f\,\in\,\mathcal M^{\theta}\,:\,[C_{\g},f]=0
$$
This algebra results to be a non commutative deformation of the algebra $A_{\g}$ of functions $\mathcal F(\R^3_{\g})$, where $\R^3_{\g}$ is the union of the orbits of the coadjoint action of the Lie algebra $\g$.
For each $\tilde A_{\g}$ we define a set of derivations in $\mathcal M_{\theta}$ which are projectable, and they have a consistent dual frame, which give the basis of 1-forms for the whole exterior algebra. In each case, we see that the differential calculus is 4d.

\section{Conclusions}
\label{sec:end}

We have considered abstract dynamical systems which in principle could represent either classical or quantum systems.
This possibility arises within the Poisson formalism on one hand or the Heisenberg formalism on the other hand, respecting the analogy principle formulated by Dirac.
Dynamical systems which are linear may be extended to the respective polynomial algebras by using Leibnitz rule and therefore they become classical or quantum only after the product rule has been specified.
This approach may turn out to be very useful if we would like to deal with mixed classical and quantum systems. These aspects will be dealt with elsewhere.

\end{document}